\documentclass[prl,amsmath,aps,floats,amssymb, floatfix, superscriptaddress, twocolumn,nofootinbib]{revtex4}

\pdfoutput=1

\usepackage{graphicx}
\usepackage{wasysym}
\usepackage{amsfonts}
\usepackage{subfigure}
\usepackage{hyperref}
\usepackage{color}        
\usepackage{ulem}

\newcommand{\sfig}[2]{
\includegraphics[width=#2]{#1}
        }
\newcommand{\Sfig}[2]{
    \begin{figure}[thbp]
    \sfig{#1.pdf}{1.1\columnwidth}
    \caption{{\small #2}}
    \label{fig:#1}
    \end{figure}
}

\newcommand{\rf}[1]{\ref{fig:#1}}

\def\lsim{\mathrel{\raise.3ex\hbox{$<$\kern-.75em\lower1ex\hbox{$\sim$}}}}
\def\gsim{\mathrel{\raise.3ex\hbox{$>$\kern-.75em\lower1ex\hbox{$\sim$}}}}

\def\cmm2{{\,\rm cm^{-2}}}
\def\cm2{{\,{\rm cm}^2}}
\def\cmm3{{\,{\rm cm}^{-3}}}
\def\gcmm3{{\,{\rm g\,cm^{-3}}}}

\def\fun#1#2{\lower3.6pt\vbox{\baselineskip0pt\lineskip.9pt
  \ialign{$\mathsurround=0pt#1\hfil##\hfil$\crcr#2\crcr\sim\crcr}}}

\def\be{\begin{equation}}
\def\ee{\end{equation}}
\def\bea{\begin{eqnarray}}
\def\eea{\end{eqnarray}}
\newcommand{\vs}{\nonumber\\}

\newcommand{\ec}[1]{Eq.~(\ref{eq:#1})}

\newcommand{\eql}[1]{\label{eq:#1}}

\begin{document}

\title{Nonlocal Gravity and Structure in the Universe}

\author{Scott Dodelson}
\affiliation{Fermilab Center for Particle Astrophysics, Fermi National Accelerator Laboratory, Batavia, IL 60510-0500}
\affiliation{Kavli Institute for Cosmological Physics, Enrico Fermi Institute, University of Chicago, Chicago, IL 60637}
\affiliation{Department of Astronomy \& Astrophysics, University of Chicago, Chicago, IL 60637}
\author{Sohyun Park}
\affiliation{Institute for Gravitation and the Cosmos, The Pennsylvania State University, University
Park, PA 16802, USA}
%\author{Sarah Shandera}
%\affiliation{Institute for Gravitation and the Cosmos, The Pennsylvania State University, University
%Park, PA 16802, USA}

\begin{abstract}
\noindent
The observed acceleration of the Universe can be explained by modifying general relativity. One such attempt is the nonlocal model of Deser and Woodard. Here we fix the background cosmology using results from the Planck satellite and examine the predictions of nonlocal gravity for the evolution of structure in the universe, confronting the model with three tests: gravitational lensing, redshift space distortions, and the estimator of gravity $E_G$. Current data favor general relativity (GR) over nonlocal gravity: fixing primordial cosmology with the best fit parameters from Planck leads to weak lensing results favoring GR by 5.9 sigma; redshift space distortions measurements of the growth rate preferring GR by 7.8 sigma; and the single measurement of $E_G$ favoring GR, but by less than 1-sigma. The significance holds up even after the parameters are allowed to vary within Planck limits. The larger lesson is that a successful modified gravity model will likely have to suppress the growth of structure compared to general relativity.
 \end{abstract}

\maketitle

\section{introduction}

The physics driving the observed acceleration of the Universe remains a mystery. Two big competing ideas for the cause are a new substance that contributes to the energy density relatively recently (dark energy) vs. a new formulation of gravity that enables acceleration even in the presence of ordinary matter~\cite{Carroll:2004de,Caldwell:2009ix,Nojiri:2010wj,Trodden:2011xa}. Models in both camps can reproduce the observed redshift-distance relation, so an emerging method to distinguish these models, to answer the (age-old) question ``Is the anomaly due to new stuff or to modified gravity?'', is to study the evolution of structure over time for a fixed expansion history~\cite{Baker:2012zs,Huterer:2013xky}. An important realization (see, e.g., \cite{Bertschinger:2006aw,Hu:2007pj,Simpson:2012ra}) is that dark energy models based on general relativity tend to predict evolution different than modified gravity models. 

Here we apply this technique to a modified gravity model proposed by Deser and Woodard~\cite{Deser:2007jk,Deser:2013uya}, designed to fit the expansion history~\cite{Koivisto:2008xfa,Deffayet:2009ca}, and recently analyzed for its effect on the growth of perturbations~\cite{Park:2012cp}.
In the last of these papers, we concluded that two measures of perturbations -- the effective Newton's constant and the gravitational slip -- differ from that in general relativity at the ten percent level. Here we study the observational implications of these deviations and confront them with current data.

We fix the redshift-distance relation and the amplitude of fluctuations, $\sigma_8$, using temperature data from the Planck satellite and polarization data from WMAP~\cite{Bennett:2012zja,Ade:2013zuv} as applied to the 5-parameter flat $\Lambda$CDM model. A given parameter set fixes the redshift-distance relation and also the initial amplitude of fluctuations but not the growth of these fluctuations. In particular, we set 
\be
\sigma_8(z_{\rm init}) = \sigma_8(z=0) \,\frac{D^{\rm GR}(z_{\rm init})}{D^{\rm GR}(0)}
,\ee
where $D$ is the growth function and the GR superscript indicates growth in general relativity. We set $z_{\rm init}=9$, long before any substantial difference between GR and nonlocal gravity arise. Once these parameters are fixed, both GR and nonlocal gravity are zero-parameter theories that make unambiguous predictions for the three tests we consider.  The comparisons are for the Planck best-fit values of the parameters: $\Omega_m=0.318, \sigma_8=0.83,h=0.67$. In the conclusion we consider the full range of allowed Planck parameters to gauge the statistical significance.

The backdrop to this test is the realization that the amplitude of the observed fluctuations in the Universe today is a bit lower than expected in the simplest $\Lambda$CDM model~\cite{Ade:2013lmv,Wyman:2013lza} given the measurement of this amplitude in the early Universe by Planck~\cite{Ade:2013zuv}. Modified gravity models then will fit the data better if they suppress growth compared to that found in a general relativity model. Nonlocal gravity does the opposite -- it enhances growth -- so is statistically disfavored.

\section{Deviations from General Relativity}
\newcommand\phib{\Phi^B}
\newcommand\psib{\Psi^B}
\newcommand\phid{\phi^D}
\newcommand\psid{\psi^D}
\newcommand\geff{G_{\rm eff}}

The perturbed Friedman-Robertson-Walker metric is written as
\be
ds^2 = -\left(1+2\Psi(t,\vec x)\right) dt^2 + a^2(t) dx^2 \left(1+2\Phi(t,\vec x)\right)\eql{metric}
.\ee
Non-relativistic particles respond to the time component of the metric, to the potential $\Psi$. In cosmology this manifests itself as a force term in the Jeans equation proportional to $\Psi$. Relativistic particles respond to both potentials, so gravitational lensing for example is governed by $\Psi-\Phi$. In general relativity, at late times, typically $\Phi=-\Psi$. 
%It is useful to connect with the notation of Bertschinger (e.g., \cite{Bertschinger:2011kk}): his potentials are related to those in \ec{metric} via
%\be
%\phib = \Psi\qquad;\qquad
%\psib = -\Phi.
%\ee
%Another useful series of papers are those by Daniel and collaborators~\cite{Daniel:2008et,Daniel:2009kr,Daniel:2010ky} whose potentials are related via
%\be
%\psid=\Psi\qquad;\qquad \phid = -\Phi
%.\ee
%This notation was also used in Ref.~\cite{Amendola:2007rr} and more recently adapted in the CFHTLens analysis~\cite{Simpson:2012ra}:
%\newcommand\geff{G_{\rm eff}}

There are several ways to parameterize deviations from general relativity. In Ref.~\cite{Park:2012cp}, we chose the convention of presenting
\bea
\eta &\equiv&  \frac{\Phi+\Psi}{\Phi}\vs
\frac{\geff}{G} &\equiv & \frac{k^2\Phi}{4\pi G\bar\rho a^2\delta}
\eql{geff}\eea
where $G$ is Newton's constant; $\bar\rho$ the mean matter density; and $\delta$ the fractional over-density in matter. Both of these functions were shown to be scale-independent in nonlocal gravity (on scales smaller than the horizon). %Fig.~\rf{fg} depicts their evolution with redshift. 
We plotted these functions using a perturbative solution, where the new source terms were evaluated using their GR values. For this work we also solved the full integro-differential equation, obtaining a more exact solution for $\eta$ and $\geff$. The solutions are similar apart from an incorrect sign in Fig.~5 of Ref.~\cite{Park:2012cp}: $\eta$ is actually negative not positive. The upshot is that, for a fixed over-density $\delta$, the potential $\vert\Psi\vert$ is larger than in GR. That is, the forcing term for the growth of structure is larger in the nonlocal model. This leads directly to the conclusion that growth is enhanced in this model.

%Two important features of these deviations from GR are that they start early, at times when canonical dark energy is not relevant and the gravitational force is weaker than in GR at early times but then becomes stronger very recently.  The first is a natural result of the fact that the nonlocal corrections to GR vanish in the radiation era and then begin to grow logarithmically after matter-radiation equality. The second is more subtle but relates to the fact that the inverse dÕAlembertian operator changes sign depending on whether it acts on a function of time or of space~\footnote{This is one of the reasons it is easy to construct a model with no observable consequences in the Solar System, as pointed out in Ref.~\cite{Deser:2013uya}.}.

%\Sfig{fg}{The two functions that quantify deviation from general relativity, as defined in \ec{geff}, as a function of redshift. At early times, the effective gravitational constant is smaller than Newton's constant so perturbations will grow slower than in general relativity (in which the plotted functions vanish).}

One way to quantify this is to follow the parameterization of Ref.~\cite{Amendola:2007rr} and more recently~\cite{Simpson:2012ra}. They defined\footnote{Note that their definition of $\Phi$ differs from ours by a minus sign.}
\bea
\Psi &\equiv& \left[ 1 + \mu(k,a) \right] \Psi_{GR}
\vs
\Psi - \Phi &=& \left[ 1+ \Sigma(k,a) \right] \,\left[\Psi_{GR} - \Phi_{GR} \right]
.\eql{defmu}\eea
%Using the facts that the two potentials are equal and opposite in GR and that the Poisson equation constrains $\Phi_{GR}/\Phi = G/\geff$ leads to:
%\bea
%\mu &=&  \frac{\geff}{G}\left(1-\eta\right) - 1
%\vs
%\Sigma &=& \frac{\geff}{G}\left(1-\frac{\eta}{2}\right) - 1
%.\eql{musig}\eea
So changes in growth of structure are determined by $\mu$ and in lensing by $\Sigma$. The relation to the first set of parameters is $1+\mu=(1-\eta)\times \geff/G$. The first factor is significantly larger than unity, overwhelming the fact that $\geff$ is slightly less than $G$, so $\mu$ is positive, and a similar conclusion hold for $\Sigma$.
Fig.~\rf{musig_vol} shows $\mu$ and $\Sigma$ as a function of redshift. Since both are positive, structure will grow faster than in general relativity and lensing will be more pronounced. 
%will be enhanced as will lensing%Also shown in Fig.~\rf{musig} is the ansatz wherein both of these functions scale as $\rho_\Lambda/(\rho_\Lambda+\rho_m)$. For the nonlocal model, this ansatz clearly fails, so the constraints obtained by the CFHTLens group~\cite{Simpson:2012ra} do not apply to this model.

\Sfig{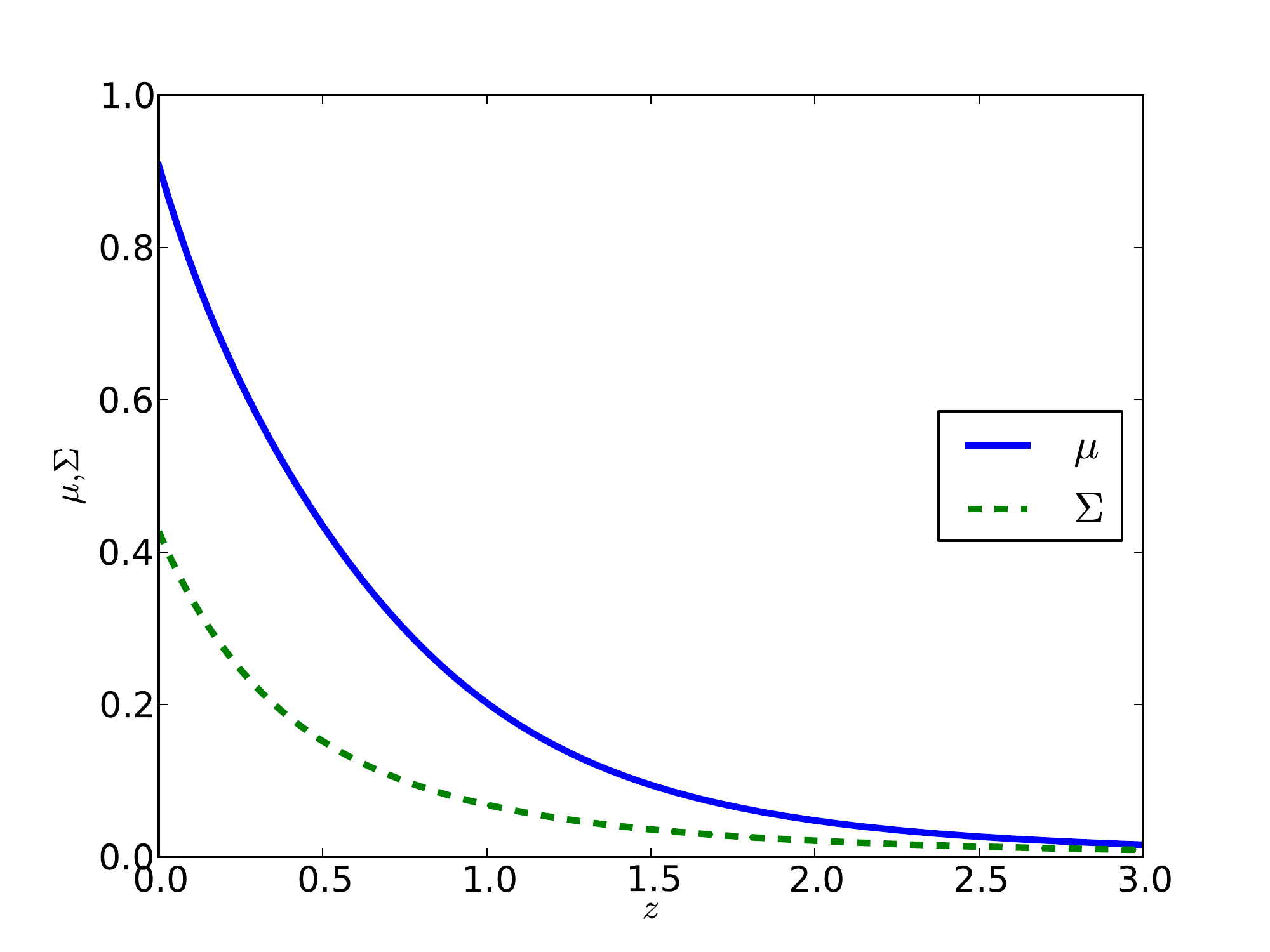}{Two functions that quantify deviations from general relativity as defined in \ec{defmu}. The growth of structure is sensitive to $\mu$ while gravitational lensing, which is determined by photon geodesics, is sensitive to $\Sigma$.}

\section{Growth of Perturbations}

The equations governing the growth of perturbations are altered in the presence of modified gravity models by the factor of $1+\mu$~\cite{Dodelson,Bertschinger:2011kk,Simpson:2012ra}
\be
\frac{d^2\delta}{da^2} + \left[ \frac{d\ln(H)}{da} + \frac{3}{a} \right]\frac{d\delta}{da}
-\frac{3}{2}\,\left[1+\mu\right]\, \frac{\Omega_m}{E^2(a)a^5}\delta = 0
\eql{gr}\ee
where $E\equiv H/H_0$. 

The {\it growth function}, $D(a)$, is the solution to \ec{gr} with initial conditions: $D(a)=a$ as is appropriate at early times ($z\sim10$) when matter dominates and over-densities simply increase with the scale factor. 
Fig.~\rf{growth_vol} shows the growth function in the nonlocal model and in $\Lambda$CDM. Starting with the same initial conditions, as measured for example by Planck, a universe governed by nonlocal gravity would become more inhomogeneous than one governed by GR. We will now see that this is not a good thing.

\Sfig{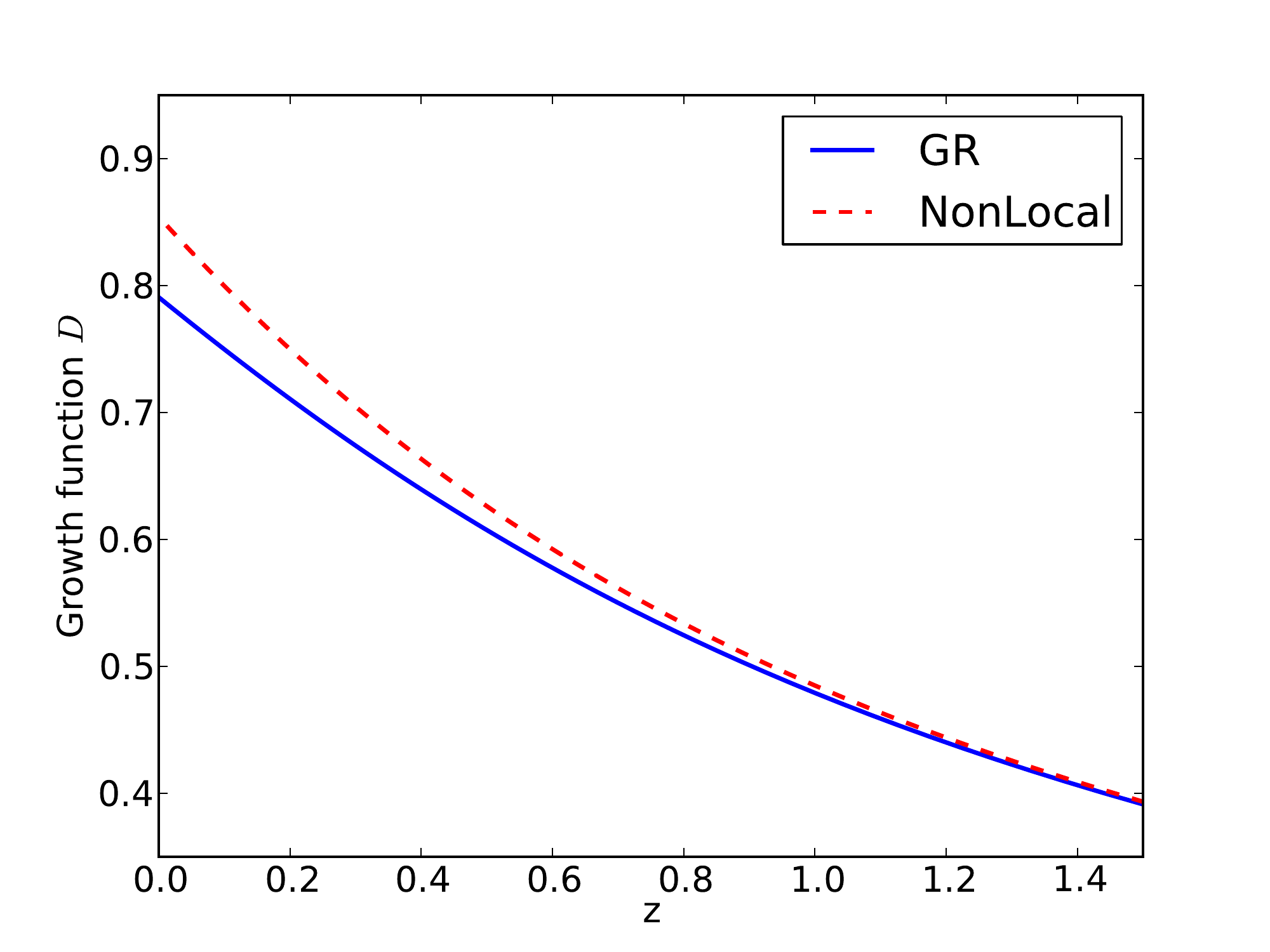}{The growth function $D(z)$ in both nonlocal gravity and general relativity with a cosmological constant.}

\section{Weak Lensing}

Lensing spectra are sensitive to both the redshift-distance relation and to the gravitational potentials. Here we fix the redshift-distance relation and the initial amplitude of fluctuations and examine the predictions for the spectrum in the presence of the modifications to gravity.  With these parameters fixed, the two models -- GR and nonlocal gravity -- have no freedom left so make unambiguous predictions. The power spectrum of the convergence of galaxies in two redshift bins is~\cite{Dodelson,Simpson:2012ra}
\bea
C_l^{ij} &=& \left(\frac{3\Omega_mH_0^2}{2}\right)^2 \,
\int_0^\infty d\chi \frac{g_i(\chi)g_j(\chi)}{a^2(\chi)}\, P(l/\chi;\chi)\vs
&&\times \left[1+\Sigma(\chi)\right]^2
,\eea
where the weighting function in each redshift bin is defined as 
\be
g_i(\chi) \equiv \int_\chi^\infty d\chi'\frac{dn_i}{d\chi'}\, \left(1-\frac{\chi}{\chi'}\right)\ee
and $dn_i/d\chi$ is the redshift distribution of source galaxies in bin $i$. The ``relativistic'' correction factor $\Sigma$ explicitly affects the spectrum, but $\mu$ also enters implicitly because of its effect on the growth and therefore on the 3D power spectrum $P$. CFHTLenS reported measurements of the angular correlation function, $\xi_+$, which can be expressed as an integral of the spectrum $C_l$ weighted by the Bessel function $J_0(l\theta)$.

Here we use the two-bin data set also used for example in Ref.~\cite{Simpson:2012ra}, as the high redshift bins are least contaminated by intrinsic alignments~\cite{2013MNRAS.432.2433H}. Fig.~\rf{xiallgr_vol} shows the auto-spectra in these two bins and the cross-spectrum along with the predictions for the nonlocal model and for standard $\Lambda$CDM.%\footnote{Note that our $\Lambda$CDM predictions agree well with those in Ref.~\cite{Simpson:2012ra}.}
%Both models use an expansion history corresponding to a flat model with $\Omega_m=0.315, h=0.67,w=-1$. The fluctuation amplitude today in $\Lambda$CDM is set to $\sigma_8=0.83$, but it is smaller in the nonlocal model because of the suppressed growth depicted in Fig.~\rf{growth}.

\Sfig{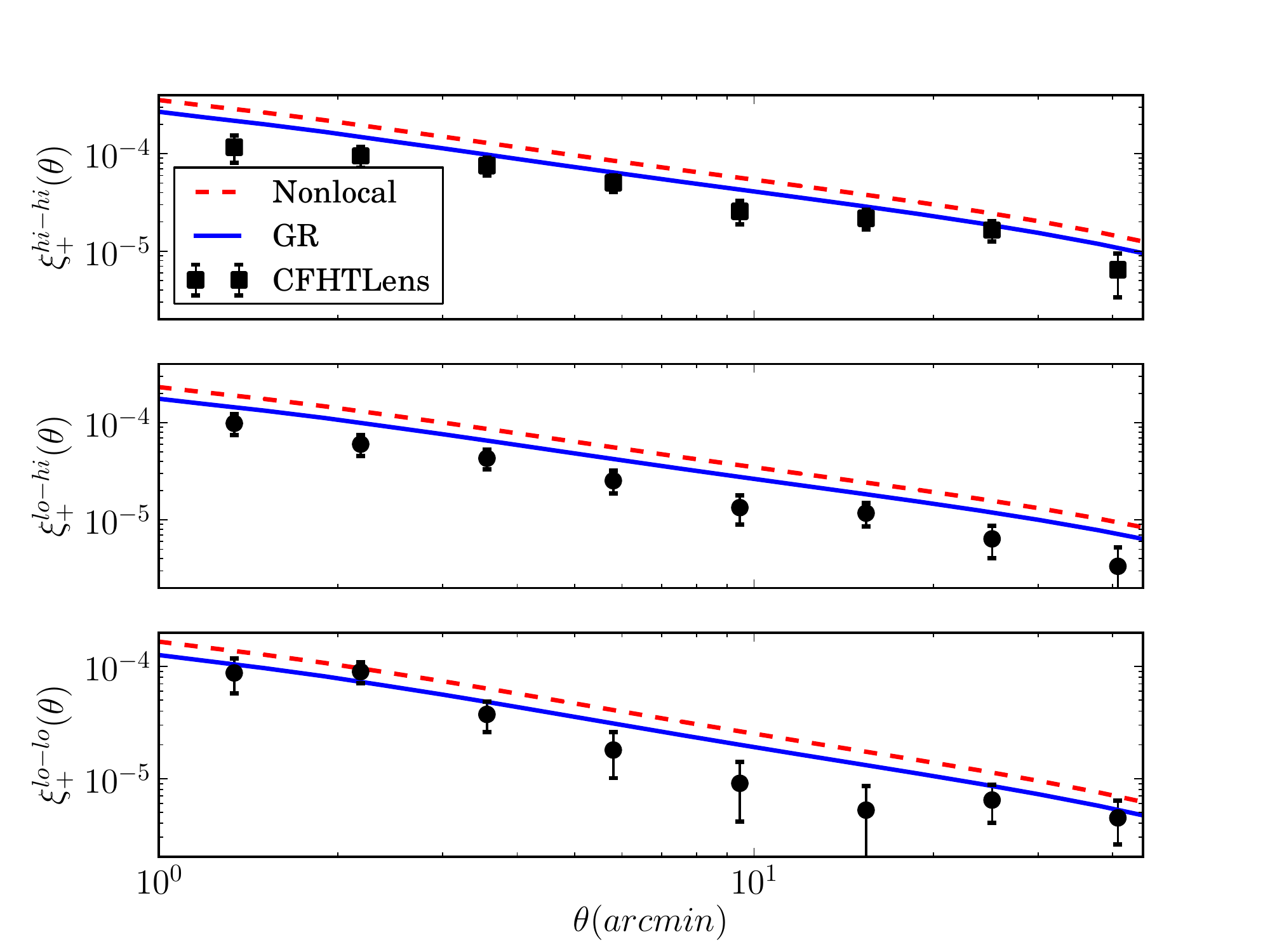}{$\xi_+$ in three different redshift bins as measured in CFHTLenS~\cite{Benjamin:2012qp} (black points with error bars). Top and bottom panels show the correlation function in the high and low redshift bins respectively, while the middle panel shows the cross spectrum. Both model have the redshift-distance relation corresponding to Planck parameters.}

The points in Fig.~\rf{xiallgr_vol} are highly correlated, especially the different spectra in the same angular bin, with correlation coefficients exceeding 0.5 at the largest angles. This reduces the statistical significance of the difference between the nonlocal model and GR.
Although the chi-by-eye in Fig.~\rf{xiallgr_vol} overwhelmingly prefers GR, including the full covariance matrix leads to a $\Delta\chi^2=34.3$ or a 5.9-$\sigma$ preference for general relativity over the nonlocal model.

\section{Redshift Space distortions}

Redshift space distortions are sensitive to the rate at which the over-densities grow.
Fig.~\rf{betas8_vol} shows the logarithmic derivative of the growth function as a function of redshift in the nonlocal model and in $\Lambda$CDM. %Differences start relatively early, hints of which can be seen in Fig.~\rf{musig_vol} and trace back to the physics of nonlocal models~\cite{Park:2012cp}. %The deviations from general relativity begin when the Ricci scalar deviates from zero, which in the cosmological context starts fairly early, after radiation domination. 
The measurements probe the product of the growth rate $\beta\equiv d\ln D/d\ln a$ and $\sigma_8(z)$, a measure of the clustering amplitude. %The curves in Fig.~\rf{betas8} were normalized so that $\sigma_8$ is fixed at early times. 

\Sfig{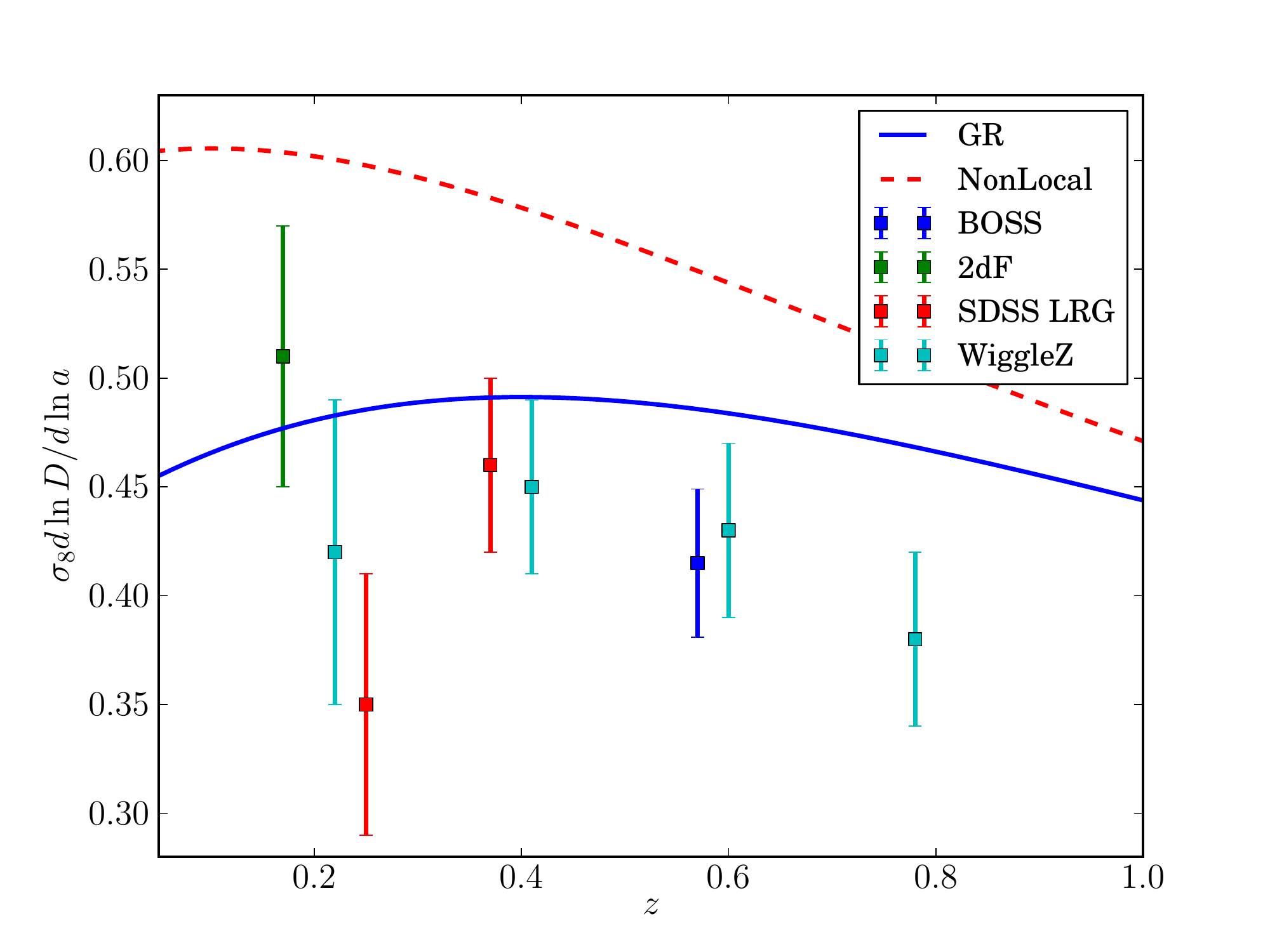}{The logarithmic derivative of the growth function as a function of redshift; this is directly measured in spectroscopic surveys capable of probing redshift space distortions. Data points come from the WiggleZ survey~\cite{Contreras:2013bol}, 2dF~\cite{2004MNRAS.353.1201P}, BOSS~\cite{Reid:2012sw} and SDSS LRG's~\cite{Samushia:2011cs}.}

A number of surveys have made measurements of $\beta\sigma_8$. Fig~\rf{betas8_vol} shows the compilation from Ref.~\cite{Samushia:2012iq} that includes various incarnations of the Sloan Survey~\cite{Samushia:2011cs,Reid:2012sw}, the 2dF galaxy survey~\cite{2004MNRAS.353.1201P}, and Wiggle-Z~\cite{Contreras:2013bol}.
These measurements are plotted against the predictions of GR and nonlocal gravity and show a 7.8-sigma preference for GR (with the parameters fixed at their best-fit values from Planck). %Upcoming surveys will improve the situation significantly. Projections for BigBOSS~\cite{Schlegel:2011zz}, as a possible implementation of the upcoming Dark Energy Spectroscopic Instrument (DESI)~\cite{Levi:2013gra}, are shown in Fig.~\rf{betas8} . A related possibility is to take low resolution spectra of a billion galaxies found in LSST, a project dubbed ``Giga-z''~\cite{Marsden:2013goa}. This would do about as well as DESI at low redshifts but extend the probe to higher redshifts, where the deviations begin to appear. Either (or both) would clearly distinguish the two classes of models at very high significance.

\section{Estimator of Gravity, $E_G$}

Gravitational lensing is sensitive to the combination $\Psi-\Phi$, while spectroscopic surveys are sensitive to the velocity field, which via the continuity equation is related to the time derivative of the over-density. Combining the two, therefore, enables~\cite{Zhang:2007nk} a test of modified gravity models. In particular, an estimator of the large scale properties of gravity is
\be
E_G \equiv \frac{k^2a(\Phi-\Psi)}{3H_0^2\beta\delta}
\ee
where $\beta\equiv d\ln D/d\ln a$ relates the velocity field to the density field. Using \ec{defmu} and the fact that $\Psi_{GR}-\Phi_{GR}=-8\pi Ga^2\bar\rho\delta/k^2$, we see that~\cite{Simpson:2012ra}
\be
E_G = \frac{\Omega_m\left[1+\Sigma\right]}{\beta}.
\ee
The growth rate $\beta$ is larger in nonlocal gravity, but $\Sigma$ is positive so there is an interesting interplay between the two effects. On balance, the $\Sigma$ enhancement wins, leading to larger values of $E_G$ in the nonlocal model.
Fig.~\rf{E_g_vol} shows this behavior as a function of redshift along with the measurement of Ref.~\cite{Reyes:2010tr}. The current measurement slightly favors GR, but upcoming measurements will do significantly better at differentiating these two models.

\Sfig{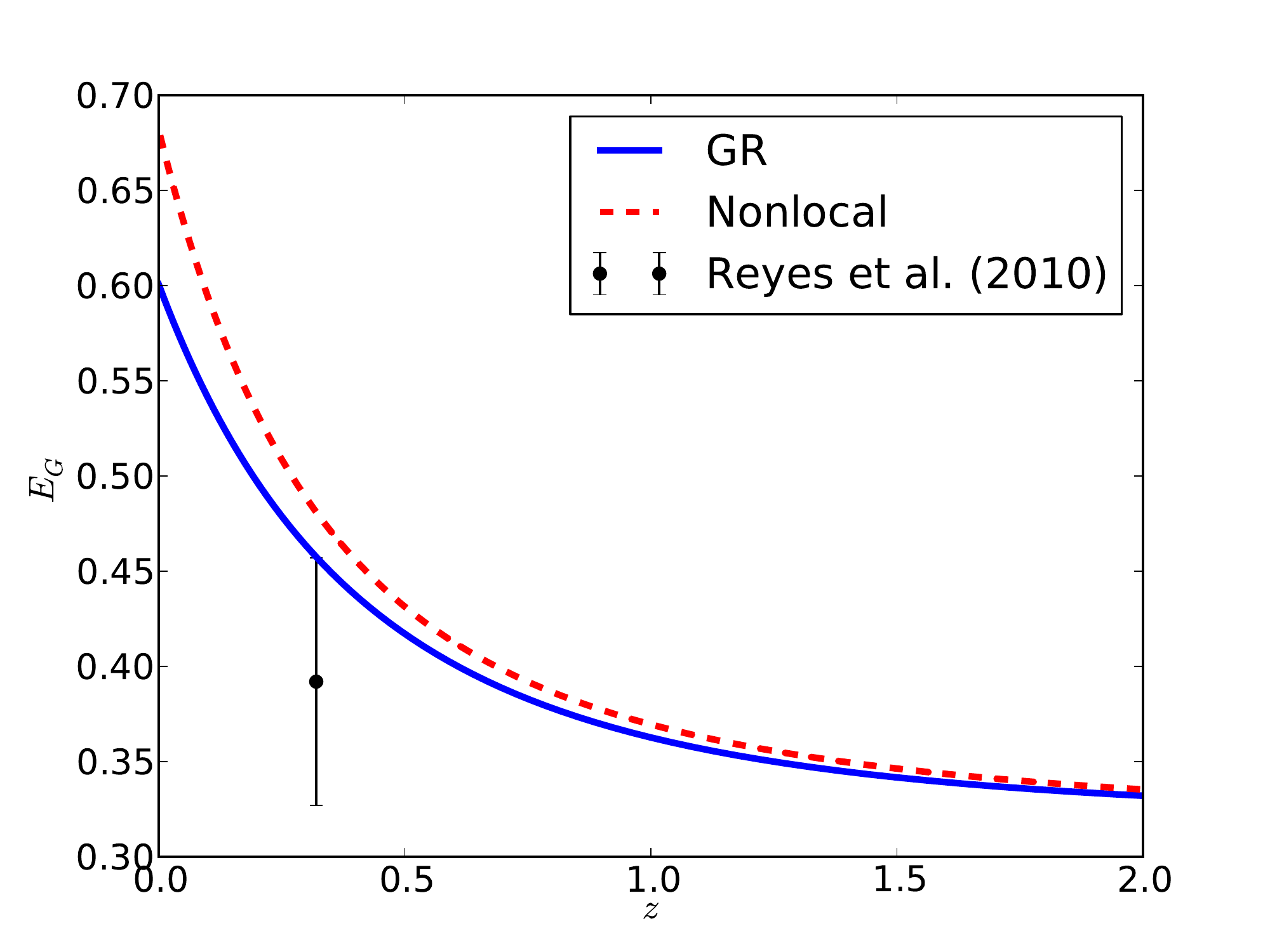}{Estimator of gravity, $E_G$, as a function of redshift in standard and modified gravity models. The data point is from Ref.~\cite{Reyes:2010tr}.}

\section{Conclusions}

Weak lensing data from CFHTLenS and measurements of the growth rate from a variety of galaxy surveys favor general relativity over nonlocal gravity with very high statistical significance. The extent to which the nonlocal model is dis-favored can be quantified in several ways. Once the redshift-distance relationship and primordial fluctuation amplitude are fixed, both models have no freedom remaining. Above we quoted the significance assuming these parameters were fixed at their best fit values as measured by Planck. This is clearly too restrictive, and relaxing this requirement lowers the statistical significance. A simple way to account for the uncertainty in the parameters is to compute the $\Delta\chi^2$ over the full region in parameter space allowed by Planck. When we do this, the mean $\Delta\chi^2$ for RSD is 58, while for lensing it is 38. A more Bayesian approach is to marginalize each likelihood (GR and nonlocal) over all parameters and compute the ratio of the marginalized likelihoods. This likelihood ratio is $8.8\times 10^7:1$, corresponding to a 6-sigma preference for general relativity.

Two other measurements of interest are the abundance of massive clusters and the Integrated Sachs-Wolfe (ISW) effect. The former likely would add to the statistical significance of the preference for GR~\cite{Ade:2013lmv}, while the latter has limited statistical power but might favor the nonlocal model, given the positive value of $\Sigma$~\cite{Fergus}. We have included neither since our calculation of the deviations breaks down on the large scales probed by the ISW effect and because we have not run simulations needed to calibrate cluster abundances.

The larger point is that the recent Universe, as measured by galaxy surveys, is relatively smooth compared to the early Universe, as measured by Planck. Many theories of modified gravity, and the nonlocal model is one of them, enhance the growth of perturbations, thereby exacerbating the conflict between the early and late Universe data. Simply put, models in which structure grows more {\it slowly} than in general relativity are currently favored. Given how difficult it has been to find a compelling alternative to general relativity, this simple clue might help guide future model building.

{\it Acknowledgments} We thank Sarah Bridle, Catherine Heymans, Beth Reid, Fabian Schmidt, Sarah Shandera, Fergus Simpson, Constantinos Skordis, Masahiro Takada, and Richard Woodard for useful suggestions and conversations. SD is
supported by the U.S. Department of Energy, including grant DE-FG02-95ER40896. SP is supported by the Eberly Research Funds of The Pennsylvania State University.

\bibliography{nonlocal}

\begin{thebibliography}{31}
\expandafter\ifx\csname natexlab\endcsname\relax\def\natexlab#1{#1}\fi
\expandafter\ifx\csname bibnamefont\endcsname\relax
  \def\bibnamefont#1{#1}\fi
\expandafter\ifx\csname bibfnamefont\endcsname\relax
  \def\bibfnamefont#1{#1}\fi
\expandafter\ifx\csname citenamefont\endcsname\relax
  \def\citenamefont#1{#1}\fi
\expandafter\ifx\csname url\endcsname\relax
  \def\url#1{\texttt{#1}}\fi
\expandafter\ifx\csname urlprefix\endcsname\relax\def\urlprefix{URL }\fi
\providecommand{\bibinfo}[2]{#2}
\providecommand{\eprint}[2][]{\url{#2}}

\bibitem[{\citenamefont{Carroll et~al.}(2005)\citenamefont{Carroll, De~Felice,
  Duvvuri, Easson, Trodden et~al.}}]{Carroll:2004de}
\bibinfo{author}{\bibfnamefont{S.~M.} \bibnamefont{Carroll}},
  \bibinfo{author}{\bibfnamefont{A.}~\bibnamefont{De~Felice}},
  \bibinfo{author}{\bibfnamefont{V.}~\bibnamefont{Duvvuri}},
  \bibinfo{author}{\bibfnamefont{D.~A.} \bibnamefont{Easson}},
  \bibinfo{author}{\bibfnamefont{M.}~\bibnamefont{Trodden}},
  \bibnamefont{et~al.}, \bibinfo{journal}{Phys.Rev.}
  \textbf{\bibinfo{volume}{D71}}, \bibinfo{pages}{063513}
  (\bibinfo{year}{2005}), \eprint{astro-ph/0410031}.

\bibitem[{\citenamefont{Caldwell and Kamionkowski}(2009)}]{Caldwell:2009ix}
\bibinfo{author}{\bibfnamefont{R.~R.} \bibnamefont{Caldwell}} \bibnamefont{and}
  \bibinfo{author}{\bibfnamefont{M.}~\bibnamefont{Kamionkowski}},
  \bibinfo{journal}{Ann.Rev.Nucl.Part.Sci.} \textbf{\bibinfo{volume}{59}},
  \bibinfo{pages}{397} (\bibinfo{year}{2009}), \eprint{0903.0866}.

\bibitem[{\citenamefont{Nojiri and Odintsov}(2011)}]{Nojiri:2010wj}
\bibinfo{author}{\bibfnamefont{S.}~\bibnamefont{Nojiri}} \bibnamefont{and}
  \bibinfo{author}{\bibfnamefont{S.~D.} \bibnamefont{Odintsov}},
  \bibinfo{journal}{Phys.Rept.} \textbf{\bibinfo{volume}{505}},
  \bibinfo{pages}{59} (\bibinfo{year}{2011}), \eprint{1011.0544}.

\bibitem[{\citenamefont{Trodden}(2011)}]{Trodden:2011xa}
\bibinfo{author}{\bibfnamefont{M.}~\bibnamefont{Trodden}},
  \bibinfo{journal}{Gen.Rel.Grav.} \textbf{\bibinfo{volume}{43}},
  \bibinfo{pages}{3367} (\bibinfo{year}{2011}), \eprint{1105.0721}.

\bibitem[{\citenamefont{Baker et~al.}(2013)\citenamefont{Baker, Ferreira, and
  Skordis}}]{Baker:2012zs}
\bibinfo{author}{\bibfnamefont{T.}~\bibnamefont{Baker}},
  \bibinfo{author}{\bibfnamefont{P.~G.} \bibnamefont{Ferreira}},
  \bibnamefont{and} \bibinfo{author}{\bibfnamefont{C.}~\bibnamefont{Skordis}},
  \bibinfo{journal}{Phys.Rev.} \textbf{\bibinfo{volume}{D87}},
  \bibinfo{pages}{024015} (\bibinfo{year}{2013}), \eprint{1209.2117}.

\bibitem[{\citenamefont{Huterer et~al.}(2013)\citenamefont{Huterer, Kirkby,
  Bean, Connolly, Dawson et~al.}}]{Huterer:2013xky}
\bibinfo{author}{\bibfnamefont{D.}~\bibnamefont{Huterer}},
  \bibinfo{author}{\bibfnamefont{D.}~\bibnamefont{Kirkby}},
  \bibinfo{author}{\bibfnamefont{R.}~\bibnamefont{Bean}},
  \bibinfo{author}{\bibfnamefont{A.}~\bibnamefont{Connolly}},
  \bibinfo{author}{\bibfnamefont{K.}~\bibnamefont{Dawson}},
  \bibnamefont{et~al.} (\bibinfo{year}{2013}), \eprint{1309.5385}.

\bibitem[{\citenamefont{Bertschinger}(2006)}]{Bertschinger:2006aw}
\bibinfo{author}{\bibfnamefont{E.}~\bibnamefont{Bertschinger}},
  \bibinfo{journal}{Astrophys.J.} \textbf{\bibinfo{volume}{648}},
  \bibinfo{pages}{797} (\bibinfo{year}{2006}), \eprint{astro-ph/0604485}.

\bibitem[{\citenamefont{Hu and Sawicki}(2007)}]{Hu:2007pj}
\bibinfo{author}{\bibfnamefont{W.}~\bibnamefont{Hu}} \bibnamefont{and}
  \bibinfo{author}{\bibfnamefont{I.}~\bibnamefont{Sawicki}},
  \bibinfo{journal}{Phys.Rev.} \textbf{\bibinfo{volume}{D76}},
  \bibinfo{pages}{104043} (\bibinfo{year}{2007}), \eprint{0708.1190}.

\bibitem[{\citenamefont{Simpson et~al.}(2012)\citenamefont{Simpson, Heymans,
  Parkinson, Blake, Kilbinger et~al.}}]{Simpson:2012ra}
\bibinfo{author}{\bibfnamefont{F.}~\bibnamefont{Simpson}},
  \bibinfo{author}{\bibfnamefont{C.}~\bibnamefont{Heymans}},
  \bibinfo{author}{\bibfnamefont{D.}~\bibnamefont{Parkinson}},
  \bibinfo{author}{\bibfnamefont{C.}~\bibnamefont{Blake}},
  \bibinfo{author}{\bibfnamefont{M.}~\bibnamefont{Kilbinger}},
  \bibnamefont{et~al.} (\bibinfo{year}{2012}), \eprint{1212.3339}.

\bibitem[{\citenamefont{Deser and Woodard}(2007)}]{Deser:2007jk}
\bibinfo{author}{\bibfnamefont{S.}~\bibnamefont{Deser}} \bibnamefont{and}
  \bibinfo{author}{\bibfnamefont{R.}~\bibnamefont{Woodard}},
  \bibinfo{journal}{Phys.Rev.Lett.} \textbf{\bibinfo{volume}{99}},
  \bibinfo{pages}{111301} (\bibinfo{year}{2007}), \eprint{0706.2151}.

\bibitem[{\citenamefont{Deser and Woodard}(2013)}]{Deser:2013uya}
\bibinfo{author}{\bibfnamefont{S.}~\bibnamefont{Deser}} \bibnamefont{and}
  \bibinfo{author}{\bibfnamefont{R.}~\bibnamefont{Woodard}}
  (\bibinfo{year}{2013}), \eprint{1307.6639}.

\bibitem[{\citenamefont{Koivisto}(2008)}]{Koivisto:2008xfa}
\bibinfo{author}{\bibfnamefont{T.}~\bibnamefont{Koivisto}},
  \bibinfo{journal}{Phys.Rev.} \textbf{\bibinfo{volume}{D77}},
  \bibinfo{pages}{123513} (\bibinfo{year}{2008}), \eprint{0803.3399}.

\bibitem[{\citenamefont{Deffayet and Woodard}(2009)}]{Deffayet:2009ca}
\bibinfo{author}{\bibfnamefont{C.}~\bibnamefont{Deffayet}} \bibnamefont{and}
  \bibinfo{author}{\bibfnamefont{R.}~\bibnamefont{Woodard}},
  \bibinfo{journal}{JCAP} \textbf{\bibinfo{volume}{0908}}, \bibinfo{pages}{023}
  (\bibinfo{year}{2009}), \eprint{0904.0961}.

\bibitem[{\citenamefont{Park and Dodelson}(2013)}]{Park:2012cp}
\bibinfo{author}{\bibfnamefont{S.}~\bibnamefont{Park}} \bibnamefont{and}
  \bibinfo{author}{\bibfnamefont{S.}~\bibnamefont{Dodelson}},
  \bibinfo{journal}{Phys.Rev.} \textbf{\bibinfo{volume}{D87}},
  \bibinfo{pages}{024003} (\bibinfo{year}{2013}), \eprint{1209.0836}.

\bibitem[{\citenamefont{Bennett et~al.}(2013)}]{Bennett:2012zja}
\bibinfo{author}{\bibfnamefont{C.}~\bibnamefont{Bennett}} \bibnamefont{et~al.}
  (\bibinfo{collaboration}{WMAP}), \bibinfo{journal}{Astrophys.J.Suppl.}
  \textbf{\bibinfo{volume}{208}}, \bibinfo{pages}{20} (\bibinfo{year}{2013}),
  \eprint{1212.5225}.

\bibitem[{\citenamefont{Ade et~al.}(2013{\natexlab{a}})}]{Ade:2013zuv}
\bibinfo{author}{\bibfnamefont{P.}~\bibnamefont{Ade}} \bibnamefont{et~al.}
  (\bibinfo{collaboration}{Planck Collaboration})
  (\bibinfo{year}{2013}{\natexlab{a}}), \eprint{1303.5076}.

\bibitem[{\citenamefont{Ade et~al.}(2013{\natexlab{b}})}]{Ade:2013lmv}
\bibinfo{author}{\bibfnamefont{P.}~\bibnamefont{Ade}} \bibnamefont{et~al.}
  (\bibinfo{collaboration}{Planck Collaboration})
  (\bibinfo{year}{2013}{\natexlab{b}}), \eprint{1303.5080}.

\bibitem[{\citenamefont{Wyman et~al.}(2013)\citenamefont{Wyman, Rudd,
  Vanderveld, and Hu}}]{Wyman:2013lza}
\bibinfo{author}{\bibfnamefont{M.}~\bibnamefont{Wyman}},
  \bibinfo{author}{\bibfnamefont{D.~H.} \bibnamefont{Rudd}},
  \bibinfo{author}{\bibfnamefont{R.~A.} \bibnamefont{Vanderveld}},
  \bibnamefont{and} \bibinfo{author}{\bibfnamefont{W.}~\bibnamefont{Hu}}
  (\bibinfo{year}{2013}), \eprint{1307.7715}.

\bibitem[{\citenamefont{Amendola et~al.}(2008)\citenamefont{Amendola, Kunz, and
  Sapone}}]{Amendola:2007rr}
\bibinfo{author}{\bibfnamefont{L.}~\bibnamefont{Amendola}},
  \bibinfo{author}{\bibfnamefont{M.}~\bibnamefont{Kunz}}, \bibnamefont{and}
  \bibinfo{author}{\bibfnamefont{D.}~\bibnamefont{Sapone}},
  \bibinfo{journal}{JCAP} \textbf{\bibinfo{volume}{0804}}, \bibinfo{pages}{013}
  (\bibinfo{year}{2008}), \eprint{0704.2421}.

\bibitem[{\citenamefont{Dodelson}(2003)}]{Dodelson}
\bibinfo{author}{\bibfnamefont{S.}~\bibnamefont{Dodelson}},
  \emph{\bibinfo{title}{Modern Cosmology}} (\bibinfo{publisher}{Academic
  Press}, \bibinfo{address}{San Diego}, \bibinfo{year}{2003}).

\bibitem[{\citenamefont{Bertschinger}(2011)}]{Bertschinger:2011kk}
\bibinfo{author}{\bibfnamefont{E.}~\bibnamefont{Bertschinger}},
  \bibinfo{journal}{Phil.Trans.Roy.Soc.Lond.} \textbf{\bibinfo{volume}{A369}},
  \bibinfo{pages}{4947} (\bibinfo{year}{2011}), \eprint{1111.4659}.

\bibitem[{\citenamefont{{Heymans} et~al.}(2013)\citenamefont{{Heymans},
  {Grocutt}, {Heavens}, {Kilbinger}, {Kitching}, {Simpson}, {Benjamin},
  {Erben}, {Hildebrandt}, {Hoekstra} et~al.}}]{2013MNRAS.432.2433H}
\bibinfo{author}{\bibfnamefont{C.}~\bibnamefont{{Heymans}}},
  \bibinfo{author}{\bibfnamefont{E.}~\bibnamefont{{Grocutt}}},
  \bibinfo{author}{\bibfnamefont{A.}~\bibnamefont{{Heavens}}},
  \bibinfo{author}{\bibfnamefont{M.}~\bibnamefont{{Kilbinger}}},
  \bibinfo{author}{\bibfnamefont{T.~D.} \bibnamefont{{Kitching}}},
  \bibinfo{author}{\bibfnamefont{F.}~\bibnamefont{{Simpson}}},
  \bibinfo{author}{\bibfnamefont{J.}~\bibnamefont{{Benjamin}}},
  \bibinfo{author}{\bibfnamefont{T.}~\bibnamefont{{Erben}}},
  \bibinfo{author}{\bibfnamefont{H.}~\bibnamefont{{Hildebrandt}}},
  \bibinfo{author}{\bibfnamefont{H.}~\bibnamefont{{Hoekstra}}},
  \bibnamefont{et~al.}, \bibinfo{journal}{MNRAS}
  \textbf{\bibinfo{volume}{432}}, \bibinfo{pages}{2433} (\bibinfo{year}{2013}),
  \eprint{1303.1808}.

\bibitem[{\citenamefont{Benjamin et~al.}(2012)\citenamefont{Benjamin,
  Van~Waerbeke, Heymans, Kilbinger, Erben et~al.}}]{Benjamin:2012qp}
\bibinfo{author}{\bibfnamefont{J.}~\bibnamefont{Benjamin}},
  \bibinfo{author}{\bibfnamefont{L.}~\bibnamefont{Van~Waerbeke}},
  \bibinfo{author}{\bibfnamefont{C.}~\bibnamefont{Heymans}},
  \bibinfo{author}{\bibfnamefont{M.}~\bibnamefont{Kilbinger}},
  \bibinfo{author}{\bibfnamefont{T.}~\bibnamefont{Erben}}, \bibnamefont{et~al.}
  (\bibinfo{year}{2012}), \eprint{1212.3327}.

\bibitem[{\citenamefont{Contreras et~al.}(2013)}]{Contreras:2013bol}
\bibinfo{author}{\bibfnamefont{C.}~\bibnamefont{Contreras}}
  \bibnamefont{et~al.} (\bibinfo{collaboration}{WiggleZ Collaboration})
  (\bibinfo{year}{2013}), \eprint{1302.5178}.

\bibitem[{\citenamefont{{Percival} et~al.}(2004)\citenamefont{{Percival},
  {Burkey}, {Heavens}, {Taylor}, {Cole}, {Peacock}, {Baugh}, {Bland-Hawthorn},
  {Bridges}, {Cannon} et~al.}}]{2004MNRAS.353.1201P}
\bibinfo{author}{\bibfnamefont{W.~J.} \bibnamefont{{Percival}}},
  \bibinfo{author}{\bibfnamefont{D.}~\bibnamefont{{Burkey}}},
  \bibinfo{author}{\bibfnamefont{A.}~\bibnamefont{{Heavens}}},
  \bibinfo{author}{\bibfnamefont{A.}~\bibnamefont{{Taylor}}},
  \bibinfo{author}{\bibfnamefont{S.}~\bibnamefont{{Cole}}},
  \bibinfo{author}{\bibfnamefont{J.~A.} \bibnamefont{{Peacock}}},
  \bibinfo{author}{\bibfnamefont{C.~M.} \bibnamefont{{Baugh}}},
  \bibinfo{author}{\bibfnamefont{J.}~\bibnamefont{{Bland-Hawthorn}}},
  \bibinfo{author}{\bibfnamefont{T.}~\bibnamefont{{Bridges}}},
  \bibinfo{author}{\bibfnamefont{R.}~\bibnamefont{{Cannon}}},
  \bibnamefont{et~al.}, \bibinfo{journal}{MNRAS}
  \textbf{\bibinfo{volume}{353}}, \bibinfo{pages}{1201} (\bibinfo{year}{2004}),
  \eprint{arXiv:astro-ph/0406513}.

\bibitem[{\citenamefont{Reid et~al.}(2012)\citenamefont{Reid, Samushia, White,
  Percival, Manera et~al.}}]{Reid:2012sw}
\bibinfo{author}{\bibfnamefont{B.~A.} \bibnamefont{Reid}},
  \bibinfo{author}{\bibfnamefont{L.}~\bibnamefont{Samushia}},
  \bibinfo{author}{\bibfnamefont{M.}~\bibnamefont{White}},
  \bibinfo{author}{\bibfnamefont{W.~J.} \bibnamefont{Percival}},
  \bibinfo{author}{\bibfnamefont{M.}~\bibnamefont{Manera}},
  \bibnamefont{et~al.} (\bibinfo{year}{2012}), \eprint{1203.6641}.

\bibitem[{\citenamefont{Samushia et~al.}(2012)\citenamefont{Samushia, Percival,
  and Raccanelli}}]{Samushia:2011cs}
\bibinfo{author}{\bibfnamefont{L.}~\bibnamefont{Samushia}},
  \bibinfo{author}{\bibfnamefont{W.~J.} \bibnamefont{Percival}},
  \bibnamefont{and}
  \bibinfo{author}{\bibfnamefont{A.}~\bibnamefont{Raccanelli}},
  \bibinfo{journal}{Mon.Not.Roy.Astron.Soc.} \textbf{\bibinfo{volume}{420}},
  \bibinfo{pages}{2102} (\bibinfo{year}{2012}), \eprint{1102.1014}.

\bibitem[{\citenamefont{Samushia et~al.}(2013)\citenamefont{Samushia, Reid,
  White, Percival, Cuesta et~al.}}]{Samushia:2012iq}
\bibinfo{author}{\bibfnamefont{L.}~\bibnamefont{Samushia}},
  \bibinfo{author}{\bibfnamefont{B.~A.} \bibnamefont{Reid}},
  \bibinfo{author}{\bibfnamefont{M.}~\bibnamefont{White}},
  \bibinfo{author}{\bibfnamefont{W.~J.} \bibnamefont{Percival}},
  \bibinfo{author}{\bibfnamefont{A.~J.} \bibnamefont{Cuesta}},
  \bibnamefont{et~al.}, \bibinfo{journal}{Mon.Not.Roy.Astron.Soc.}
  \textbf{\bibinfo{volume}{429}}, \bibinfo{pages}{1514} (\bibinfo{year}{2013}),
  \eprint{1206.5309}.

\bibitem[{\citenamefont{Zhang et~al.}(2007)\citenamefont{Zhang, Liguori, Bean,
  and Dodelson}}]{Zhang:2007nk}
\bibinfo{author}{\bibfnamefont{P.}~\bibnamefont{Zhang}},
  \bibinfo{author}{\bibfnamefont{M.}~\bibnamefont{Liguori}},
  \bibinfo{author}{\bibfnamefont{R.}~\bibnamefont{Bean}}, \bibnamefont{and}
  \bibinfo{author}{\bibfnamefont{S.}~\bibnamefont{Dodelson}},
  \bibinfo{journal}{Phys.Rev.Lett.} \textbf{\bibinfo{volume}{99}},
  \bibinfo{pages}{141302} (\bibinfo{year}{2007}), \eprint{0704.1932}.

\bibitem[{\citenamefont{Reyes et~al.}(2010)\citenamefont{Reyes, Mandelbaum,
  Seljak, Baldauf, Gunn et~al.}}]{Reyes:2010tr}
\bibinfo{author}{\bibfnamefont{R.}~\bibnamefont{Reyes}},
  \bibinfo{author}{\bibfnamefont{R.}~\bibnamefont{Mandelbaum}},
  \bibinfo{author}{\bibfnamefont{U.}~\bibnamefont{Seljak}},
  \bibinfo{author}{\bibfnamefont{T.}~\bibnamefont{Baldauf}},
  \bibinfo{author}{\bibfnamefont{J.~E.} \bibnamefont{Gunn}},
  \bibnamefont{et~al.}, \bibinfo{journal}{Nature}
  \textbf{\bibinfo{volume}{464}}, \bibinfo{pages}{256} (\bibinfo{year}{2010}),
  \eprint{1003.2185}.

\bibitem[{\citenamefont{Simpson}(2013)}]{Fergus}
\bibinfo{author}{\bibfnamefont{F.}~\bibnamefont{Simpson}},
  \bibinfo{journal}{Private communication}  (\bibinfo{year}{2013}).

\end{thebibliography}
\end{document}